\documentclass[superscriptaddress,longbibliography,floatfix,twocolumn,nobibnotes]{revtex4-2}
\usepackage{upgreek}
\usepackage{amsmath,braket}
\usepackage{graphicx}
\usepackage{mathtools}
\usepackage{color}
\usepackage{setspace}
\newcommand{\RNum}[1]{\uppercase\expandafter{\romannumeral #1\relax}}
\usepackage{soul}
\usepackage{hyperref}
\usepackage{enumitem}
\usepackage{graphicx}
\usepackage{hyperref}
\usepackage{multirow}   
\usepackage{tabularx}   
\usepackage{makecell}

\begin{document}
\title{A Concise Primer on Solid-State Quantum Emitters}

\author{Shicheng Yu}
\thanks{These two authors contributed equally}
\affiliation{Institute of Fundamental and Frontier Sciences, University of Electronic Science and Technology of China, Chengdu, 610054, P. R. China}
\affiliation{Tianfu Jiangxi Laboratory, Chengdu, 641419, P. R. China}
\author{Xiaojie Zhang}
\thanks{These two authors contributed equally}
\affiliation{Institute of Fundamental and Frontier Sciences, University of Electronic Science and Technology of China, Chengdu, 610054, P. R. China}
\affiliation{Tianfu Jiangxi Laboratory, Chengdu, 641419, P. R. China}
\author{Xia Lei}
\affiliation{Institute of Fundamental and Frontier Sciences, University of Electronic Science and Technology of China, Chengdu, 610054, P. R. China}
\affiliation{Tianfu Jiangxi Laboratory, Chengdu, 641419, P. R. China}
\author{Liang Zhai}
\email{liang.zhai@uestc.edu.cn}
\affiliation{Institute of Fundamental and Frontier Sciences, University of Electronic Science and Technology of China, Chengdu, 610054, P. R. China}
\affiliation{Tianfu Jiangxi Laboratory, Chengdu, 641419, P. R. China}

\begin{abstract}
{Quantum emitters serve as essential on-demand photonic resources, generating quantum states of light such as single photons and entangled photon pairs while serving as interfaces between light and matter. Buried in the solid state, quantum emitters enable a straightforward adoption of advanced nanofabrication techniques, facilitating precise engineering of their photonic environment for scalable quantum technologies. In this review, we introduce the fundamentals of quantum emitters and the key metrics characterising their performance. We highlight three material platforms: quantum dots, defect centres in diamond, and defect centres in silicon carbide. We summarise the recent developments of these platforms and discuss their advancements in quantum applications, including quantum communication, computation, and sensing. Finally, we provide a comparison across the three platforms, along with an outlook on future directions and potential challenges.
}
\end{abstract}

\maketitle
\normalsize

\section{Introduction} \label{sec:intro} 

Quantum applications, including quantum communication, quantum information processing, and quantum sensing, have gained widespread attention due to their potential to surpass classical counterparts. These applications require quantum information to be encoded, transmitted, stored, and processed, necessitating a variety of devices with distinct functions. Quantum emitters, typically nanoscale objects with discrete energy levels, play a crucial role by enabling single-photon generation, acting as spin-based quantum memories, and facilitating coherent spin-photon interfaces, making them integral to quantum technologies.

Amid rapid developments in quantum technologies, numerous quantum emitters have been studied. While at different stages of development, it is timely to examine their properties against key metrics for quantum applications. We highlight three emitter platforms in solids: epitaxial quantum dots, defect centres in diamond, and defect centres in silicon carbide (see Fig.\ \ref{fig:fig1}). Quantum dots (QDs) are bright and fast quantum light sources capable of producing desired photonic states on demand. In comparison, defect centres in diamond and silicon carbide (SiC) offer strong potential for spin-photon interfaces, exhibiting relatively long coherence times even at room temperature. Other emerging platforms, such as defects in two-dimensional materials and silicon, also show promise but remain outside the scope of this review. 

We begin this review by introducing the key metrics relevant to quantum applications in Section \ref{sec:chap2}. Section \ref{sec:chap3} then presents the material platforms, covering the emitter fabrication, energy-level structures, quantum optical properties, and integration capabilities. We discuss in Section \ref{sec:chap4} the quantum applications involving the three emitter platforms, focusing on recent progress in quantum communication, computation, and sensing. Finally, as an outlook, we compare these platforms in Section \ref{sec:chap5}, highlighting their respective strengths, challenges, and future research directions.
\vspace{-10pt} 

\section{Basics of quantum emitters}\label{sec:chap2}

Quantum emitters refer to a unique class of nanoscale structures that, upon optical or electrical triggering, create a non-classical state of light. A fundamental example of quantum emitters is atoms. Due to their discrete energy levels dictated by the quantisation of electron energies in the atomic potential, atoms can emit single photons or entangled photon pairs through well-defined energy transitions\cite{Wilk2007}. However, atoms (or ions) require precise control over their spatial positioning, which is achieved by trapping them in optical or magnetic fields\cite{Pinkse2000, Barry2014}, thereby minimising thermal motion and enabling stable, well-isolated quantum states. This cooling process introduces significant experimental overhead.

Quantum emitters are also found in solid-state systems\cite{Lodahl2015, Aharonovich2016, Lohrmann2017, Heindel2023, Zhou2023}, utilising either material defects or artificially engineered nanoscale structures like quantum dots or wires. Unlike atoms, solid-state emitters eliminate the need for complex trapping techniques, offering a notable advantage. Additionally, their dipole moments are often orders of magnitude larger than those of atoms, as the physical dimensions of solid-state emitters typically span thousands of crystal lattices. This leads to significantly faster photon emission rates, which is highly desirable for light-emitting applications. Moreover, by leveraging advanced nanofabrication technologies, solid-state emitters can be integrated into various photonic nanostructures, enabling scalable and integrable device designs. These combined advantages make solid-state emitters a popular choice for quantum light sources. 

However, there are notable challenges. Unlike atomic systems, where uniformity is typically maintained when the species are the same, solid-state emitters often exhibit large variability. This leads to a wide spread in emission properties, such as wavelength and lifetime. Another critical factor is the quality of the crystal or material in which the emitters reside, as lattice imperfections directly impact the emitted photon characteristics. Ensuring high material purity and minimising defects is crucial for achieving high-quality quantum light sources.

The quality of quantum emitters is characterised by several key metrics. Below, we outline these metrics, highlighting their importance in evaluating emitter performance and suitability for quantum applications.
\begin{itemize}[leftmargin=*]    
    \item \textbf{Radiative rate:} This metric is inversely linked to the emitter's radiative lifetime, indicating the speed at which the emitter creates photons. For a bright emitter, the radiative rate should be high, typically on the order of hundreds of MHz or GHz. 
    
    \item \textbf{Spectra and linewidth:} This metric characterises the quantum emitter in the spectral domain. Noise during photon emission -- such as the thermal, charge, and spin noise \cite{Kuhlmann2014, Zhai2022} -- directly impacts the emitter's performance. From a thermal noise perspective, the coherent part of the emission spectrum is referred to as the zero-phonon line (ZPL), representing the transition that occurs without phonons' involvement. This line is accompanied by a broad, incoherent phonon sideband\cite{Iles2017}. The \textit{ZPL portion} within the emission spectrum indicates the fraction of ``useful" photons, whereas the ZPL \textit{linewidth} reveals information about other noise processes. Ideally, the ZPL portion should approach unity, and the linewidth should be narrow. However, there is a fundamental limit to how narrow the linewidth can be: in the absence of decoherence, the minimal linewidth is the inverse of the radiative lifetime, known as the Fourier limit or transform limit. The ratio between the measured linewidth and this limit serves as a key parameter for assessing the impact of decoherence on the emitter \cite{Trusheim2020, Zhai2020}.
    
    \item \textbf{Quantum efficiency:} Quantum efficiency (intrinsic) refers to the ratio of the radiative rate to the total decay rate (comprising both radiative and non-radiative rates). For an ideal quantum emitter, the photon creation process (i.e., excited-state preparation and radiative relaxation) should be \textit{deterministic}, requiring a quantum efficiency of unity.
    
    \begin{figure*}[t!]
    \centering
    \includegraphics[width = 14 cm]{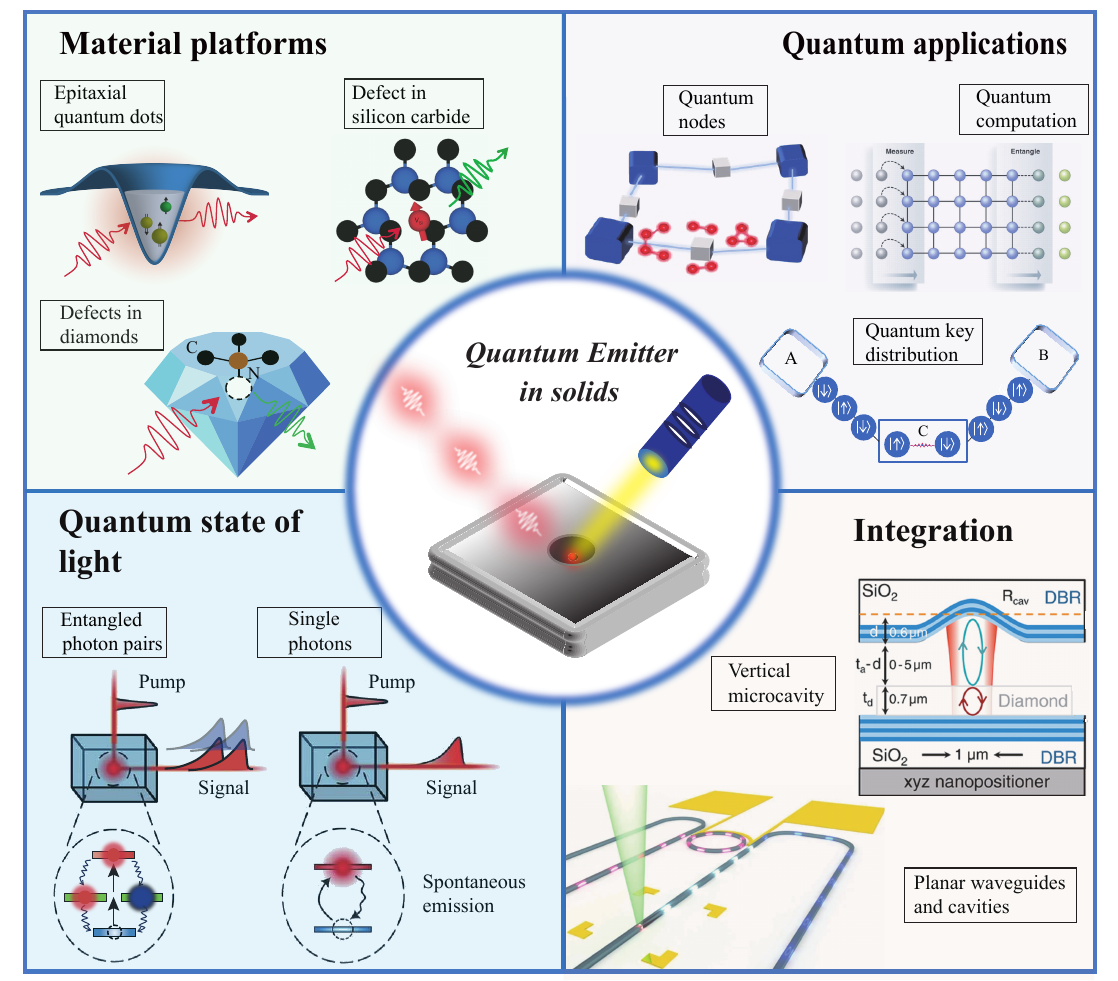}
    \caption{An overview of quantum emitters, encompassing material platforms, quantum photonic resource generation, potential applications, and prospects for integration and scalability. Part of the figure is reproduced with permission from: top left, Ref.\ \cite{Zhai2022} and Ref.\ \cite{Dietz2023}, under a Creative Commons license CC BY 4.0; bottom left, Ref. \cite{Esmann2024}, under a Creative Commons license CC BY 4.0; bottom right, Ref.\ \cite{Flagan2022} and Ref.\ \cite{Elshaari2017}, under a Creative Commons license CC BY 4.0; and top right, Ref.\ \cite{Obrien2007}, AAAS. } 
    \label{fig:fig1}
    \end{figure*}

    \item \textbf{Single-photon purity:} This metric is quantified by the second-order correlation function at zero time delay, $g^{(2)}(0)$, measured in a Hanbury Brown and Twiss (HBT) experiment\cite{Brown1956}(see Fig.\ \ref{fig:fig2}a). The purity is defined as 1 - $g^{(2)}(0)$, and is particularly important for single-photon sources. Ideally, a single-photon emitter should produce no more than one photon at a time, corresponding to $g^{(2)}(0) = 0$.

    \item \textbf{Indistinguishability:} Indistinguishability reflects the coherence of the generated photons and determines whether they are quantum-mechanically identical in properties such as frequency, polarisation, and emission time. It is evaluated via Hong-Ou-Mandel (HOM) interference\cite{Hong1987}: when two identical photons interfere at a beamsplitter, they bunch and exit the same output port, whereas completely distinct photons have a 50\% chance of exiting opposite outputs. Ideally, indistinguishable photons exhibit unit interference contrast\cite{braNczyk2017} (see Fig.\ \ref{fig:fig2}b).
    
    \item \textbf{Entanglement fidelity:} This is a key metric for entangled photon sources, quantifying how closely the generated state matches the ideal entangled state. High fidelity indicates minimal decoherence and noise, while lower fidelity points to imperfections in the entangled state. A common approach for characterising entanglement fidelity is quantum state tomography\cite{James2001}, as illustrated in Fig.\ \ref{fig:fig2}c for polarisation entanglement. However, quantum state tomography scales poorly with the qubit number. For larger systems, fidelity witnesses offer a practical alternative: GHZ states are assessed via parity measurements in the computational basis and correlation measurements in rotated bases\cite{Otfried2007,Li2020,Meng2024}, while cluster states are verified using stabiliser measurements\cite{Otfried2007, Li2020, Abobeih2022}.
    
    \item \textbf{Brightness:} While quantum efficiency determines whether photons can be created deterministically, brightness relates to the efficiency with which these photons are collected from the emitter. Ideally, each trigger applied to the emitter should produce exactly one photon. A suboptimal collection efficiency reduces this probability, leading to erasure errors (loss of quantum information) in quantum technologies. Emitters embedded in solids inherently suffer from low collection efficiency due to the high refractive index of the substrate materials, which causes total internal reflection. One way to mitigate this issue is to engineer the local photonic environment of the emitter, for example, by coupling the emitter's dipole to a single cavity mode or waveguide mode\cite{Tomm2021, Herrmann2024}.
    
    \item \textbf{Light-matter interface:} For quantum emitters to operate as nodes in quantum networks, they must not only possess a photonic interface but also a matter qubit capable of processing and temporarily storing quantum information\cite{Kimble2008}. Strong light-matter coupling is crucial to facilitate efficient exchange between the two. Similar to its photonic counterpart, high-fidelity control of the matter qubit is essential. Additionally, the matter qubit must have a sufficiently long \textit{coherence time}, characterised by the $T_2^*$ and $T_2$ values, which are typically obtained from Ramsey and Hahn echo sequences, respectively (see Fig.\ \ref{fig:fig2}d for the Hahn echo; for Ramsey, omit the central $\pi$ pulse)\cite{Becker2018,Wolfowicz2021}.

    \item \textbf{Scalability:} Scalability involves several aspects. First, the quantum emitter must be compatible with nanophotonic chips that enable active, programmable photon routing functionalities. This requires either monolithic or hybrid integration\cite{Babin2022,Zhou2023,Ding2024}, along with compatibility for nanofabrication processes. Additionally, integration with optical fibres, such as through fibre pigtailing and stable packaging\cite{Bremer2022}, is beneficial. Second, multiple emitters, rather than just one, must be operable in quantum applications, which presents challenges related to the uniformity and fine-tuning of the quantum emitters. Third, it is also crucial that in practical applications the emitter’s emission wavelengths fall within the telecom bands\cite{Yu2023}, specifically the O-band and C-band, where losses are minimised.
\end{itemize}

\begin{figure*}[ht!]
    \centering
    \includegraphics[width = 15 cm]{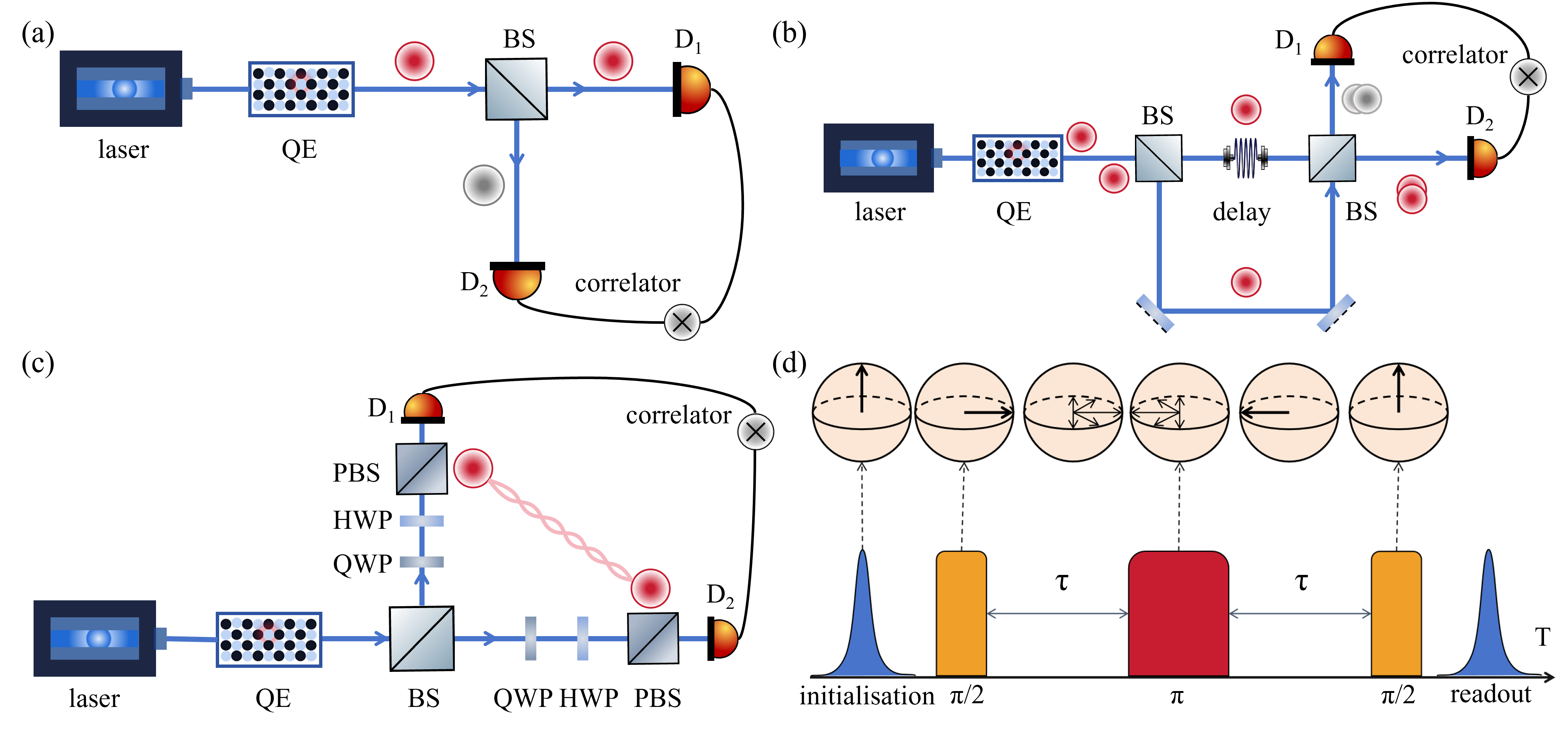}
    \caption{\textbf{Characterisation of quantum emitters.} 
    \textbf{(a) Hanbury Brown and Twiss experiment:} Photons generated by a quantum emitter (QE) are directed to a 50:50 beamsplitter (BS) and detected at two separate detectors for coincidence counting. Since a true single photon cannot be split, the $g^{(2)}(0)$ should ideally approach zero. 
    \textbf{(b) Hong-Ou-Mandel experiment:} Photons emitted by the QE pass through an unbalanced Mach-Zehnder interferometer and arrive simultaneously at the second BS. When the photons are quantum mechanically indistinguishable, quantum interference ensures that they always exit the same output port, leading to a characteristic dip in coincidence measurements.
    \textbf{(c) Quantum-state tomography (polarisation encoding):} Entangled photons from the QE pass through polarisation optics and are measured in different computational bases via coincidence counting, enabling density matrix reconstruction and entanglement fidelity extraction.
    \textbf{(d) Hahn echo sequence:} After initialisation, a $\pi/2$ pulse flips the qubit to the equator of the Bloch sphere. A waiting time $\tau$ follows, after which a $\pi$-pulse is applied, refocusing the qubit at the same interval $\tau$. A final $\pi/2$ pulse projects the qubit back into the computational basis for readout. This sequence effectively mitigates slowly varying noise, acting as a high-pass filter.
    }
    \label{fig:fig2}
\end{figure*}
Quantum emitters must be prepared in specific excited states to generate the desired photonic states. A common approach is photoluminescence\cite{Hogele2008}, where a laser with energy higher than the emitter's transition energy is used for excitation. In QDs, this typically involves the generation of electron-hole pairs in higher-energy bands, which then relax to the lowest exciton state before radiative recombination\cite{Lodahl2015}. For defect centres in insulators, electrons are promoted to higher vibrational levels before relaxing to the lowest vibrational state\cite{Du2024}. Although photoluminescence is a simple and efficient method for characterising the emitter's transition spectrum, relaxation from higher energy levels introduces timing jitter, resulting in reduced photon coherence\cite{Borri2005}.

To mitigate time jitter, excitons must be created directly at the transition energy levels. This requires an excitation laser with energy comparable to or exactly matching the transition energy of the emitter (i.e., the ZPL energy), a technique known as \textit{resonant excitation}. Notably, when using a pulsed laser with a pulse width shorter than the emitter’s decoherence and dephasing times, a two-level system can be coherently driven to the excited state with near-unit fidelity\cite{zhai2019_2, Fischer2017}.

The scattered light from the excitation laser must be efficiently separated from the generated photons to ensure that only the latter are collected at the output. Various methods have been developed to eliminate the residual laser, including utilising the polarisation difference between the excitation laser and the generated photons\cite{Zhai2020_2}, employing a three-level system\cite{Wei2022, Sbresny2022, Yan2022, Zhai2022a}, using the assistance of an optical cavity\cite{Wang2019b, Thomas2021, Wu2023} and applying spectral\cite{He2019, Koong2021, Karli2022} or temporal filtering\cite{Humphreys2018}.

\section{Materials platforms for quantum emitters}
\label{sec:chap3}
This chapter provides an overview of quantum emitter platforms, highlighting the material properties most pertinent to their use in quantum photonic applications. A comparative summary is presented in Table~\ref{table:table1}.

\subsection{Epitaxial Quantum Dots}

With their exceptional photonic properties, epitaxial (self-assembled) QDs stand out as one of the most advanced material platforms for realising photonic quantum resources. \\

\begin{figure*}[ht!]
    \centering
    \includegraphics[width = 18 cm]{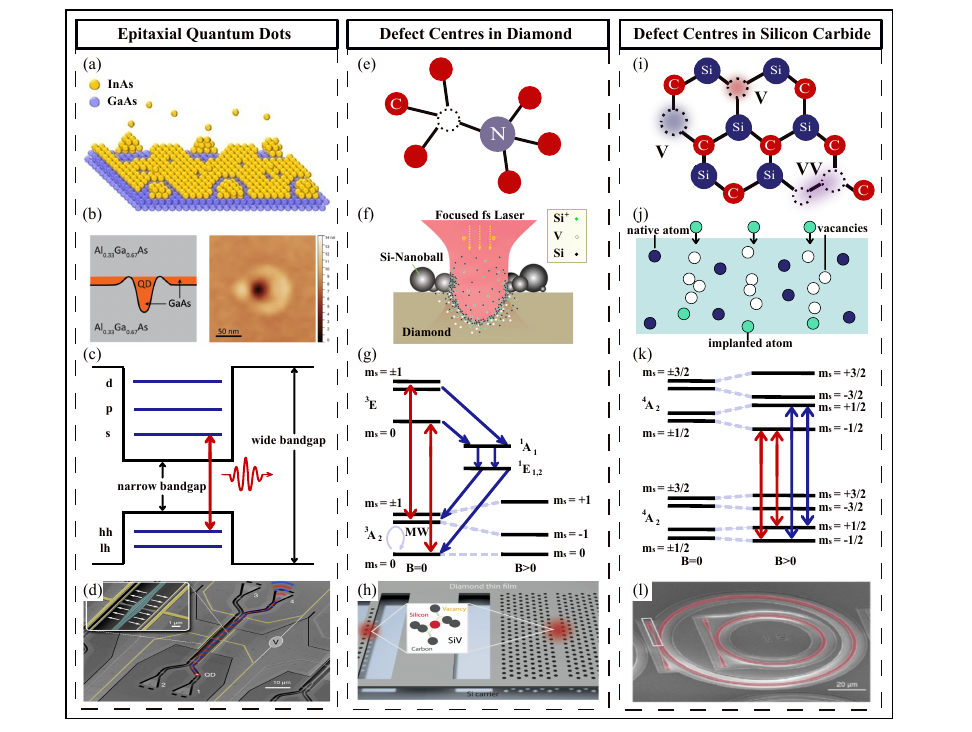}
    \caption{\textbf{Material Platforms for quantum emitters.}
    (a) Sketch of InAs nano-islands grown on GaAs.
(b) Illustration of GaAs QDs prepared by local droplet etching.
(c) Diagram of the QD's level structures.
(d) False-coloured image of a nano-electromechanical switch fabricated on a suspended GaAs membrane. 
(e) Sketch of the atomic structure of a nitrogen-vacancy defect centre in diamond.
(f) Sketch illustrating the laser-writing technique for preparing SiV$^-$ defect centres in diamond.
(g) Energy level structure of NV$^-$ in diamond.
(h) Thin-film diamond membrane fabricated into 1D and 2D photonic crystal cavities with embedded silicon vacancies.
(i) Sketch of the atomic structure of defect centres in SiC.
(j) Illustration of the ion implantation method for preparing defect centres in solids.
(k) Energy level structure of the silicon vacancy in SiC.
(l) Microring resonators fabricated on a SiCOI platform. Panels reproduced with permission from:
(a) Ref.\ \cite{Wang2020a}; (b) Ref.\ \cite{chen2024}; (d) Ref.\ \cite{Chanana2022}; (f) Ref.\ \cite{Ju2021}; (h) Ref.\ \cite{Ding2024}; (l) Ref.\ \cite{Guidry2020}, under a Creative Commons license CC BY 4.0.
    }
    \label{fig:fig4s}
\end{figure*}

\noindent\textbf{Quantum Dot Growth}\\
\noindent Epitaxial QDs are typically fabricated from III-V compound semiconductors using molecular beam epitaxy (MBE) or metal-organic chemical vapour deposition (MOCVD). These growth techniques, conducted in an ultrahigh vacuum environment, ensure the growth of highly pure materials with precise control over deposition parameters. Among epitaxial growth methods, the Stranski-Krastanov (SK) mode has been developed and refined over decades to produce high-optical-quality QDs. By leveraging the lattice mismatch between the QDs and the surrounding matrix material, this method induces strain, which triggers a transition from layer-by-layer growth to nanoisland nucleation (Fig.\ \ref{fig:fig4s}a). Subsequent capping of the nanoislands yields optically active QDs.

While the SK mode is widely adopted, alternative methods such as droplet epitaxy (DE)\cite{Heyn2007,Gurioli2019} and local droplet etching (LDE)\cite{Babin2021,Gurioli2019} overcome the constraints of lattice mismatch, enabling a broader range of material combinations\cite{Gurioli2019}. Droplet epitaxy utilises metal droplets that self-assemble and crystallise on the substrate surface, forming island-like nanostructures. In contrast, LDE employs high-temperature annealing to initiate a nanohole drilling process. The elevated annealing temperature reduces impurities and defects. The etched nanoholes are subsequently filled with quantum-dot materials and capped with the substrate (Fig.\ \ref{fig:fig4s}b). Both DE and LDE methods facilitate the formation of QDs with enhanced symmetry\cite{Huo2013}, owing to significantly reduced intrinsic strain. This symmetry is advantageous for applications such as entangled photon sources, as will be elaborated on later.  

Despite the availability of well-established growth techniques, the QD growth continues to be a highly active field. For instance, a recent study \cite{Xin2024} investigates the growth of III-V and II-V QDs on van der Waals surfaces, highlighting emerging strategies for versatile material integration and near-surface optical interfaces.\\

\noindent\textbf{Level Structures}\\
\noindent Due to their nanoscale dimensions, QDs exhibit discrete energy levels in both the conduction and valence bands, see Fig.\ \ref{fig:fig4s}c. In the conduction band, the levels are labelled as s-, p-, d-shells, etc., similar to those in atoms. In the valence band, the orbital degree of freedom is coupled with spin, and the corresponding band structures (top of the valence band) are typically divided into heavy-hole (hh) and light-hole (lh), based on their effective mass. In most cases, the heavy-hole band lies above the light-hole band\cite{Testelin2009}. A recently established method for precisely determining the QD band structures involves the radiative Auger effect\cite{Lobl2020, Spinnler2021}. This technique has been used to map out the shell structures in the conduction band, as well as the various hole states in the valence band\cite{Yan2024}.

An s-shell electron bounded with a heavy hole forms the fundamental excitonic state in QDs, the neutral exciton $|\textrm{X}^{0}\rangle$. Upon spontaneous emission, this excitonic state relaxes to its ground state $|\textrm{g}\rangle$, i.e., an empty QD state, and emits a photon. In the ideal case, $|\textrm{X}^{0}\rangle-|\textrm{g}\rangle$ is a two-level system. However, due to exchange interaction arising from the symmetry breaking of the QD\cite{Bayer2002}, the $|\textrm{X}^{0}\rangle$ state splits into two distinct levels, $|\textrm{X}_{\textrm{a}}\rangle$ and $|\textrm{X}_{\textrm{b}}\rangle$, which are separated by a small energy difference known as the fine-structure splitting (FSS). Apart from neutrally charged states, QDs can also trap electrons and holes, forming charged excitonic states\cite{Zhai2020} such as $|\textrm{X}^{-}\rangle$, $|\textrm{X}^{+}\rangle$, etc. $|\textrm{X}^{0}\rangle$, $|\textrm{X}^{-}\rangle$, and $|\textrm{X}^{+}\rangle$ are all optically active under resonant excitation, whereas other charged states are typically not -- the relaxation of these states is forbidden by the selection rule\cite{Lodahl2015}. Besides, QDs also host the biexciton state $|\textrm{XX}\rangle$\cite{chen2024, Chen2016}, where the energy $E_{\textrm{XX}} = 2 E_{\textrm{X}} - E_{\textrm{B}}$, with $E_{\textrm{B}}$ being the binding energy. The states $|\textrm{XX}\rangle$, $|\textrm{X}\rangle$, and $|\textrm{g}\rangle$ forms a three-level cascade. When the fine structure splitting is vanishingly small (FSS $\approx$ 0), the two decay channels from $|\textrm{XX}\rangle$ to $|\textrm{g}\rangle$ via $|\textrm{X}_{\textrm{a}}\rangle$ and $|\textrm{X}_{\textrm{b}}\rangle$, respectively, are symmetric. Under these conditions, a maximally entangled photon pair can be generated, expressed as $|\Psi\rangle \propto |\sigma^+_{\textrm{XX}}\rangle|\sigma^-_{\textrm{X}}\rangle+|\sigma^-_{\textrm{XX}}\rangle|\sigma^+_{\textrm{X}}\rangle$, where $\sigma^{+/-}$ denotes circular polarisation.\\

\noindent\textbf{Optical Properties}\\
\noindent 
The emission wavelengths of epitaxial QDs vary, with GaAs/AlGaAs QDs (hereafter denoted as GaAs QDs) emitting at around 780 nm\cite{Zhai2020_2, Huber2017}, and InGaAs/GaAs QDs (hereafter InGaAs QDs) emitting at around 920 nm\cite{Liu2018, Uppu2020}. Recent advancements have extended the emission of InGaAs QDs to the telecom O-band and C-band, approximately 1310 nm and 1550 nm, respectively. This has been achieved through approaches, such as the introduction of strain-relief buffers in the SK mode\cite{Zeuner2021, Chen2017}, or by droplet etching\cite{Spitzer2024}. Moreover, InAs/InP QDs can also emit single photons in the telecom C-band \cite{Yu2023, Vajner2024}. The emission wavelengths of QDs can be tuned by adjusting their size by exploiting the fine control of growth parameters, enabling modulation across a certain range. \\

From a spectral perspective, QDs typically exhibit a ZPL comprising approximately 95\% of the emission, along with near-unity intrinsic quantum efficiency due to negligible non-radiative decay\cite{Somaschi2016}. Their radiative rate generally falls within 1 to 4 GHz, corresponding to an optimal linewidth in the hundreds of MHz range (Fourier limit). The measured linewidth can approach the Fourier limit when the noise processes accompanying the emission are controlled. An effective way to mitigate noise in QDs is by embedding them into a charge-tunnelling device, such as a p-i-n diode\cite{Kuhlmann2014,Zhai2020}. In such devices, the QD energy level can be tuned via an external voltage relative to the Fermi level in the doped region, and the charge state is deterministically locked by the Coulomb blockade\cite{Kuhlmann2014}, reducing charge noise. This technique has been successfully implemented in both InGaAs and GaAs QDs, with GaAs QD linewidths reduced to only about 8\% above the Fourier limit \cite{Zhai2020}.

Quantum dots can function as two-level systems and be excited using a pulsed resonant laser -- this yields an excellent \textit{deterministic} single-photon source. The measured single-photon purity is largely limited by experimental imperfections (such as residual component from the excitation laser), and can achieve values as high as $1-g^{(2)}(0)>99\%$\cite{Scholl2019,Schweickert2018,Hanschke2018}. The coherence of the generated photons, quantified through the HOM interference, is dictated by noise. Sequentially produced photons are generally identical, as only the ``fast" noise is relevant. Photons separated by a longer interval are usually less coherent due to the influence of noise with a broader bandwidth. This effect was clearly demonstrated in experiments, where indistinguishability drops from 98\% at a 13 ns photon interval to 92\% at a 2 $\mu$s interval\cite{Wang2016, Thoma2016}. However, with refined semiconductor engineering, noise can be substantially suppressed. In low-noise GaAs QDs\cite{Zhai2020}, indistinguishability remains at 98\% even for microsecond intervals and only drops to a minimal value of 93\% at near-infinite intervals\cite{Zhai2022}.

Quantum dots can also work as entangled photon sources when excited to the biexciton state. Decay from the three-level cascade emits polarisation-entangled photon pairs, when FSS is minimised. The FSS is related to the morphology of the QD and can be tuned to near-zero by either externally applied strain\cite{Chen2016,Huber2018,Ou2022}, electric field\cite{Zhang2017}, or light field\cite{chen2024}. Utilising multi-axial strain, sub-microelectronvolt FSS value and entanglement fidelity of 98\% has been reported\cite{Huber2018}. The polarisation-entangled photon state can also be conveniently converted into path or time-bin entanglement by employing the chiral optical effect\cite{Ostfeldt2022, Siampour2023} or an unbalanced interferometer setup\cite{Prilmuller2018}, respectively.

The brightness of QD-based photon sources is among the highest of all quantum emitters. Collection efficiency is enhanced by various micro- or nano-cavities\cite{Fischbach2017,Liu2018,Schweickert2018,Wang2019b,Tomm2021,Wu2023,Maring2024}, as well as single-mode waveguides\cite{Uppu2020}. By employing an open-cavity design\cite{Tomm2021}, the \textit{end-to-end efficiency} of a single-photon source reaches 71.2\%\cite{Ding2025} under resonant pulsed excitation. This efficiency is defined as the ratio between the repetition rate of the excitation laser and the single-photon count rate at the output fibre\cite{Tomm2021}. For example, when the QD is excited by an 80 MHz repetition-rate laser, the single-photon output is around 57 MHz.

Quantum dots can trap a single electron or hole as a spin qubit under an external magnetic field. Together with a charged trion, they form two three-level systems with shared ground states (refer to Ref.\ \onlinecite{Warburton2013} for the exact level structures). High-fidelity spin control has been demonstrated via optical pulses, either through a virtual two-photon Raman process\cite{Bodey2019} or ultrafast laser excitation via the AC Stark effect\cite{Zaporski2023}. This control capability has enabled the creation of spin-photon entanglement, with fidelities exceeding 80\%\cite{DeGreve2012, Coste2023} and entanglement generation rates reaching the MHz regime\cite{Coste2023}. Electron spin coherence is limited by random nuclear spin fluctuations, with intrinsic $T_2^*$ times typically a few nanoseconds. By cooling the nuclear spins with an algorithmic feedback scheme, the $T_2^*$ can be extended to 125 ns for InGaAs QDs\cite{Jackson2022} and 0.6 $\mu$s for GaAs QDs\cite{Nguyen2023}. The electron-spin $T_2$ times for InGaAs and GaAs QDs are typically several microseconds, measured by a Hahn echo. While in InGaAs QDs the spin coherence is governed by a strained environment\cite{Stockill2016}, leading to inhomogeneous electric field gradients that limit the further extension of $T_2$, dynamic decoupling (DD) sequences can extend $T_2$ in unstrained GaAs QDs to a sub-millisecond (0.11 ms) regime\cite{Zaporski2023}. \\

\noindent\textbf{Integration}\\
\noindent 
III-V semiconductors, such as GaAs and AlGaAs, exhibit strong nonlinear and electro-optic effects\cite{Chen2024c}. Their compatibility with standard nanofabrication enables low-loss waveguides and high-Q cavities \cite{Uppu2020, Papon2019}, with quality factors nearing $10^6$ in open micro-cavities and around $10^5$ in planar photonic crystal cavities. In both cases, these high-Q factors were achieved through surface passivation\cite{Najer2019}. For scalable integration, precise alignment of nanophotonic structures to QDs is essential. Various positioning methods have been developed, including site-controlled growth of QDs\cite{Zhang2018, Drosse2020}, micro-PL imaging for pre-fabrication positioning\cite{Liu2024a}, and in-situ electron beam lithography\cite{Schnauber2018}.

As a monolithic platform, QDs embedded in nanophotonic waveguides or cavities form deterministic photon-emitter interfaces. This interface enables on-chip creation of quantum photonic states\cite{Uppu2020} and provides nonlinear functionality\cite{Javadi2015, Thyrrestrup2017, Sun2018} for single-photon switches and transistors. Beyond the nonlinearity provided by the interface itself, opto-electromechanical devices have also been implemented on GaAs chips (see Fig.\ \ref{fig:fig4s}d) to actively switch and route single photons from QD emission\cite{Papon2019}, achieving a 23 dB splitting ratio with sub-microsecond response times. Notwithstanding, the monolithic platform can suffer from relatively high losses (e.g., compared to silicon nitride), particularly when embedding QDs in a vertical p-i-n diode to suppress charge noise, as doping increases the scattering of light. This hinders large-scale applications. To overcome this, heterogeneous integration into mature photonic platforms has been explored, enabling fast, reconfigurable photon routing. Hybrid integration with material platforms such as silicon-on-oxide\cite{Kim2017}, silicon nitride\cite{Chanana2022,Pholsen2025}, lithium niobate\cite{Aghaeimeibodi2018}, and silicon carbide\cite{Zhu2022} has been tested, utilising pick-and-place or transfer printing techniques, achieving coupling efficiencies above 50\%\cite{Zhu2022,Pholsen2025}. 


\subsection{Defect centres in diamond}

Atomic vacancies (defects) in diamond serve as an alternative type of quantum emitter, associated with spin qubits that have long coherence times. We restrict our discussion to single-crystal diamond here. \\

\noindent\textbf{Creation of defect centres in diamond}\\
\noindent The creation of defect centres should begin with high-quality diamond materials and minimise crystal damage. Depending on the type of atomic defect, these can include nitrogen vacancies (NV centres) or group IV vacancies, such as silicon (SiV centres), germanium (GeV centres), or Tin (SnV centres). Their formation typically involves incorporating the desired atomic species (Si, Ge, Sn, etc.) into the diamond and displacing carbon atoms from their lattice sites to create vacancies (Fig.\ \ref{fig:fig4s}e). A common method is ion implantation \cite{Sipahigil2016, Schroder2017, Wan2020}, followed by annealing and overgrowth\cite{Rugar2020}. Ion implantation via focused ion beam (FIB) allows precise lateral control of defect positioning and tunability of defect depth\cite{Machielse2019}. Subsequent high-temperature annealing mobilises the dopants and vacancies, forming the desired defect centres\cite{Toyli2010} while also mitigating implantation-induced damage, thereby improving the quality of the defects. Additionally, high-pressure environments suppress the diamond-to-graphite transition, allowing annealing at higher temperatures -- known as the HPHT annealing method -- which has been shown to enhance defect quality further\cite{Zhang2023, Tang2024}.

An alternative method involves laser writing, particularly for the creation of NV centres. Unlike group IV defect centres, NV centres have a non-vanishing permanent electric dipole moment, making them susceptible to surface noise and charge fluctuations in the surrounding environment\cite{Chakravarthi2021}. The ion implantation method introduces damage that diminishes the quality of NV centres, particularly when these defects are created in thin membranes or nanostructures\cite{Ruf2019, Yurgens2021}. In contrast, laser writing involves irradiating a femtosecond pulse laser onto the diamond substrate\cite{Chen2019, Ju2021}, causing minimal damage to the crystal while avoiding implantation and resulting in NV centres with reduced charge noise\cite{Yurgens2021}. Additionally, the laser writing technique can be generalised to group IV defect centres; for instance, SiV centres have been created by irradiating diamond coated with a thin layer of Si nanoballs\cite{Rong2019} (see Fig.\ \ref{fig:fig4s}f).\\

\noindent\textbf{Level Structures}\\
\noindent 
Nitrogen vacancies in diamond usually exist in two charge states: the neutral state NV$^0$ and the negatively charged state NV$^-$. In the emission spectrum, the transition energies for NV$^-$ and NV$^0$ lie around 1.95 eV and 2.15 eV, respectively. The NV$^-$ state is preferred in quantum applications, as it provides a higher fluorescence intensity\cite{Santori2010} and a more stable spin. 

The energy levels of NV$^-$ consist of a $^{\textrm{3}}\textrm{A}$ ground state and a $^{\textrm{3}}\textrm{E}$ excited state, both are spin triplets, each with the sublevels $\textrm{m}_\textrm{s} = 0$ and $\textrm{m}_\textrm{s} = \pm 1$. Dipolar interaction between the electron spins splits the $\textrm{m}_\textrm{s} = \pm 1$ doublet and the $\textrm{m}_\textrm{s} = 0$ level by approximately 2.87 GHz and 1.41 GHz (zero-field splitting) for the ground and excited states, respectively, in the absence of magnetic and strain fields\cite{Fuchs2008}. Under an external magnetic field, the degeneracy of the $\textrm{m}_\textrm{s} = \pm 1$ doublet is lifted by the Zeeman effect, with splitting proportional to the field strength, see Fig.\ \ref{fig:fig4s}g. Spin-preserved transitions occur between the excited and ground triplet states, a radiative process. In addition, the excited state can relax with a small probability via intersystem crossing to the singlet states, $^{\textrm{1}}\textrm{A}$ and $^{\textrm{1}}\textrm{E}$, which is mostly a non-radiative process.

Group IV colour centres feature distinct energy levels compared to their nitrogen counterparts. For instance, SiV$^-$ has both ground and excited states that are orbital- and spin-doublets. The orbital degeneracy is lifted by spin-orbit coupling and the Jahn-Teller effect, resulting in four possible optical transitions, each with double spin degeneracy (s = 1/2, unlike the s = 1 in NV$^-$). In low-strain environments, the ground- and excited-state splittings are $\sim$50 GHz and $\sim$250 GHz, respectively\cite{Muller2014b}. The small ground-state splitting leads to phonon-mediated dephasing, limiting the operation to millikelvin temperatures \cite{Becker2018}. Heavier group IV vacancies, such as GeV$^-$ and SnV$^-$ exhibit larger ground-state splittings ($\sim$170 GHz and $\sim$890 GHz), enabling higher-temperature operation. To mitigate phonon-induced dephasing, strategies such as strain \cite{Sohn2018,Guo2023} and phononic engineering \cite{Kuruma2022} can be applied. Strain not only increases ground-state splitting but also induces orbital mixing, allowing efficient microwave control of the spin ground states in GeV$^-$ and SnV$^-$. Phononic structures, such as phononic crystals, reduce thermal phonon density by introducing a phononic bandgap \cite{Kuruma2022}.\\

\noindent\textbf{Optical Properties}\\
\noindent 
The emission wavelengths of defect centres in diamond range from approximately 600 nm to 700 nm (see Table\ \ref{table:table1}). Despite their relatively high intrinsic quantum efficiency, NV centres exhibit a ZPL that accounts for only $\sim$3\% of the total emission at around 4K \cite{Santori2010}, necessitating extensive filtering to preserve photon coherence. In contrast, group IV centres offer significantly higher Debye–Waller factors, with ZPL contributions of up to $\sim$70\%, $\sim$60\% and $\sim$60\% for SiV$^-$, GeV$^-$ and SnV$^-$ centres\cite{Tzeng2024, Siyushev2017, Gorlitz2020}.

The radiative rates of these defect centres are slower than QDs, but still relatively fast, with NV$^-$, SiV$^-$, GeV$^-$ and SnV$^-$ approximately radiating at a rate of $2\pi \times (13,\ 90,\ 28,\ 32)$ MHz, respectively\cite{Trusheim2014, Sipahigil2014, Bhaskar2017, Rugar2020_2}. Their linewidths are measured at cryogenic temperatures for noise characterisation using photoluminescence excitation (PLE), where a laser resonantly excites the ZPL transition and the phonon sideband emission is collected. This method is preferred over resonant fluorescence due to the sub-unity Debye-Waller factor (especially for NV centres) and the ease of removing the excitation laser. The measured linewidths approach the transform limit: for NV$^-$, 14 MHz versus a 12.4 MHz limit \cite{Chen2017_2}, achieved via laser writing at a 50 µm depth. Reported linewidths for SiV$^-$, GeV$^-$, and SnV$^-$ are 100 MHz, 42 MHz, and 36 MHz, respectively\cite{Siyushev2017, Zuber2023, Rugar2020_2}. However, scan duration strongly affects these values — shorter scans often yield narrower linewidths, while integration over minutes reveals broader features due to slow noise sources like spectral diffusion. For group IV centres, linewidths can broaden to several hundred MHz over long acquisitions. Minimising spectral diffusion and charge instability, especially in shallow emitters embedded in membranes or nanophotonic devices, remains a key challenge.

Emissions from diamond defect centres demonstrate good single-photon purity, with \(1 - g^{(2)}(0)\) values typically ranging from 97\% to 99\% under continuous-wave excitation\cite{Wang2022, Chen2022}. Pulsed single-photon generation has also been demonstrated by integrating emitters into nanophotonic structures to enhance ZPL emission, achieving raw purities of 90–95\%\cite{Arjona2022, Bersin2024}. However, generating highly indistinguishable photons remains challenging due to low ZPL contribution, charge noise, and spectral instability. Narrow spectral filtering and temporal post-selection help improve indistinguishability but reduce photon count rates. Near 90\% indistinguishability has been achieved for NV and SiV centres\cite{Pompili2021, Bersin2024}, lower values have been reported for GeVs (60\%)\cite{Chen2022} and SnVs (63\%)\cite{Arjona2022}. Additionally, HOM interference between two remote NV centres has demonstrated 79\% visibility\cite{Stolk2022}.

Due to the longer radiative lifetime and smaller Debye-Waller factors, defect centres in diamond are typically less bright compared to emitters like QDs. The brightness of the ZPL photons is enhanced when these centres are incorporated into micro- or nanocavities. A saturated count rate above 100 kHz has been achieved using NV centres embedded in an open microcavity \cite{Yurgens2024}, with continuous-wave excitation and recording of the ZPL photoluminescence count. Similar designs hosting SnV$^-$ centres have demonstrated a count rate of approximately 15 kHz\cite{Herrmann2024}, leaving room for improvements.

The defect vacancies in diamond exhibit excellent spin properties due to their wide bandgap, sparse nuclear-spin environment, weak spin-orbit interactions, and the ability to grow high-purity materials \cite{Atature2018}. The spins of defect centres are typically initialised and read out optically, while being rotated through microwave drives. High-fidelity spin control has been achieved\cite{Pingault2017}, although the rotation speeds typically lie within the MHz range. In addition to microwave control, spin rotation using ultrafast pulses \cite{Becker2016} and surface acoustic waves \cite{Maity2020} have also been developed. With these spin control capabilities, along with the long spin coherence times detailed below, defect centres in diamond can serve as entanglement sources for both spin and photon. Entanglement fidelities above 90\% have been achieved \cite{Stas2022}. Furthermore, the electron spin can mediate interactions between the photon and a nearby nucleus --an operation known as the PHONE gate -- creating photon-nucleus entanglement, currently with an attainable fidelity of $\sim$85\%\cite{Stas2022}.

At room temperature, NV$^-$ centres in isotopically purified diamond have demonstrated Ramsey $T_2^*$ times surpassing 1.5 ms and Hahn echo $T_2$ times beyond 2 ms\cite{Herbschleb2019}. Lowering temperatures further extends the coherence, with $T_2$ exceeding 1 s at liquid helium temperatures\cite{Abobeih2018} using dynamical decoupling. Group IV centres generally exhibit shorter spin coherence times\cite{Sukachev2017, Trusheim2020, Guo2023}. Ramsey $T_2^*$ values are typically on the order of a few microseconds \cite{Sohn2018, Senkalla2024}. The $T_2$ times under dynamical decoupling can reach 13 ms for SiV$^-$ at 100 mK \cite{Sukachev2017}, 20 ms for GeV$^-$ at millikelvin temperatures \cite{Senkalla2024}, and 1.6 ms for SnV$^-$ in strained membranes at ~2 K \cite{Guo2023}.\\

\noindent\textbf{Integrations}\\
\noindent 
Diamond is among the hardest and most inert materials found in nature, making nanofabrication on diamond particularly challenging. Albeit difficulties, significant progress has been made in integrating defect centres into diamond micro- and nanophotonic cavities, such as photonic crystal cavities \cite{Ding2024, Bersin2024}(Fig.\ \ref{fig:fig4s}h), microrings\cite{Faraon2011, Hausmann2014}, microdisks\cite{Shandilya2022}, and open microcavities\cite{Reindl2017, Yurgens2024}. In microstructures (e.g., microrings and microcavities), quality factors as high as $1\times10^6$ have been achieved\cite{Hausmann2014}. These structures typically rely on relatively thick diamond membranes, often exceeding one micron. On the one hand, this thickness benefits NV centres by reducing the influence of surface noise for defects buried at greater depths; on the other hand, it imposes constraints on miniaturisation and integration. By utilising thin-film diamond a few hundred nanometres thick, this quality factor has recently been extended to the order of $10^5$ in photonic crystal cavities\cite{Ding2024}. State-of-the-art nanofabrication on diamond is exemplified in Ref.\ \onlinecite{Kuruma2024}, where a phononic crystal structure of $\sim70$ nm thickness with $\sim20$ nm tethers and a radius of curvature has been fabricated using quasi-isotropic etching.

Diamond must also be integrated with other platforms to enable active photon routing. Recent efforts have demonstrated a pick-and-place approach that transferred 16 micro-chiplets, each containing 8 prefabricated diamond waveguides hosting SiV or GeV centres, onto an integrated aluminium nitride platform\cite{Wan2020}, as well as the large-scale integration of 1,024 waveguide channels onto a CMOS chip \cite{Li2024}. Direct bonding has also been employed, allowing thin-film diamond membranes to be integrated with silicon, fused silica, sapphire, and lithium niobate, where photonic nanostructures can be patterned post-integration\cite{Guo2024}.

\subsection{Defect centres in silicon carbide}

Silicon carbide is a wide-bandgap semiconductor (bandgap: 2.3-3.2 eV) with diverse polytypes, including hexagonal 4H–SiC, 6H–SiC, and cubic 3C–SiC. Similar to diamond (bandgap: $\sim$ 5.5 eV), silicon carbide hosts a variety of optically active atomic defect centres that interface single photons and a spin qubit.\\

\noindent\textbf{Creation of defect centres in silicon carbide}\\
\noindent 
Similar to diamond, the creation of defect centres in SiC begins with the preparation of high-quality substrate materials, which are usually either bulk SiC or SiC membranes on insulators, achieved through high-temperature chemical vapour deposition\cite{Kimoto2001}. A broad range of defect centres exist in SiC (Fig.\ \ref{fig:fig4s}i), including neutral divacancies $\textrm{V}_{\textrm{Si}}\textrm{V}_{\textrm{C}}^{\ \textrm{(0)}}$, negative silicon vacancies $\textrm{V}_{\textrm{Si}}^{\textrm{\ (-)}}$, nitrogen-vacancy pairs $\textrm{V}_{\textrm{Si}}\textrm{N}_{\textrm{C}}^{\ \textrm{(-)}}$, among many others. For a detailed summary of these defects, readers are referred to prior works\cite{Lohrmann2017,Atature2018,Castelletto2020}. \\

Here, we focus on two well-studied vacancy types, $\textrm{V}_{\textrm{Si}}\textrm{V}_{\textrm{C}}^{\ \textrm{(0)}}$ and $\textrm{V}_{\textrm{Si}}^{\textrm{\ (-)}}$, in hexagonal SiC. Both defect centres can be created using methods similar to those employed for diamond, including electron-beam irradiation \cite{Christle2014}, ion implantation \cite{Falk2013, Wang2017c, Hu2024} (Fig.\ \ref{fig:fig4s}j), and direct laser writing \cite{Chen2019b,Day2023}, albeit with tailored procedures. Techniques such as focused ion beam and laser writing enable precise control over defect number, lateral position, and depth, facilitating the creation of arrays of localised defect centres suitable for subsequent integration with photonic nanostructures \cite{Day2023}.\\

\noindent\textbf{Level Structures}\\
\noindent 
The silicon vacancies in SiC are associated with different configurations, known as V$1$, V$1^\prime$ and V$2$, while divacancies generate ZPL lines labelled from PL$1$ to PL$7$, corresponding to different vacancy sites (PL$1$ -- PL$4$) and other modifications\cite{Widmann2015, Nagy2019, Castelletto2020, He2024}. The lattice configuration of SiC polytypes influences the defects' properties, causing shifts in emission energy and altering spin behaviour\cite{Soltamov2015}. The energy levels of the V$1$, V$2$, and divacancy defects resemble those of the NV centres in diamond. As shown in Fig.\ \ref{fig:fig4s}k, the silicon vacancies in SiC compromise an orbital singlet spin quartet ground state, and a similar excited state (a spin-3/2 system). Dipolar spin-spin interactions cause a splitting between the ground-state $\textrm{m}_\textrm{s} = \pm1/2$ and $\textrm{m}_\textrm{s} = \pm3/2$ states, a zero-field splitting. The ground-state splitting for the V$1$ centre is approximately 5 MHz \cite{Nagy2019}, while for V$2$, it is around 70 MHz \cite{Widmann2015}. Divacancies are spin-1 defects, with the ground state split between $\textrm{m}_\textrm{s} = \pm1$ and $\textrm{m}_\textrm{s} = 0$, along with a small splitting between $\textrm{m}_\textrm{s} = \pm1$ due to zero-field interaction \cite{Miao2019}. Both silicon vacancies and divacancies enable spin-selective optical transitions under resonant excitation \cite{Nagy2019}. In addition, there are non-radiative pathways via intersystem crossings, albeit complicated level configurations. \\

\noindent\textbf{Opticlal Properties}\\
\noindent 
SiC defect centres exhibit a wide spread of emission wavelengths, spanning from around 600 nm to the telecom O-band \cite{Castelletto2020, Wang2018_2}. Among these defects, V1 centres in 4H-SiC emit at around 862 nm (ZPL), V2 at around 917 nm, and divacancies at around 1100 nm. Both silicon vacancies and divacancies are accompanied by a broad and significant phonon-sideband. V1 centres feature a relatively large ZPL portion, with up to 40\% observed for an isolated defect\cite{Nagy2018}, albeit in a nanostructure and ignoring the excited-state V1$^\prime$ line. V2 centres and divacancies are typically weaker in intensity, with ZPL accounting for less than 10\% of the total emission. 

Protected by symmetry, the emission from silicon vacancies is relatively stable and exhibits a linewidth close to the Fourier limit. Under resonant excitation, the linewidth of a V1 centre was measured to be around 60 MHz, about twice its inverse lifetime \cite{Nagy2019} (5.5 ns lifetime, transform limit around 29 MHz). For the V2 centre, linewidths as narrow as $\sim 40$ MHz have been observed, also about twice the lifetime limit\cite{Vandestolpe2024}. However, severe spectral diffusion broadens this linewidth to the GHz regime over longer timescales. On the other hand, divacancy centres exhibit broader linewidths, typically in the range of 100–200 MHz in 4H-SiC, accompanied by a longer lifetime, which corresponds to a transform limit of approximately 11 MHz\cite{Christle2017}. Similar to diamond, the linewidth broadening is primarily due to spectral diffusion induced by charge noise. When integrated into a p-i-n device, a 31 MHz linewidth was recorded on a divacancy over a 3-hour PLE scan, indicating a stable charge environment\cite{Anderson2019_2}. However, the mechanism here is different from that of QDs -- charge depletion, rather than Coulomb blockade, mitigates the charge noise\cite{Anderson2019_2}. 

The purity of the single photon created by defects in SiC can reach up to $1 - g^{(2)}(0) \sim 95\%$ under continuous-wave excitation\cite{Christle2017,Li2021}, and approximately 88\% under pulsed excitation\cite{Morioka2020}, with both methods employing off-resonant excitation due to relatively weak signals. With a narrow spectral filter of 29 MHz and temporal gating, an indistinguishability value of nearly 90\% was recorded\cite{Morioka2020} for photons separated by a $\sim$ 50 ns interval. The brightness of the source, however, remains relatively low, ranging from tens to hundreds of kilohertz in saturation for silicon vacancies or divacancies\cite{Li2021}, and reaching up to MHz levels for certain specific defects\cite{Castelletto2014} — all measured under continuous-wave excitation. The collection efficiency must be improved, for example, by integrating into nano-cavities or waveguides, while the non-radiative decay channels of the defects should be suppressed through, e.g., charge state engineering.

Like NV centres in diamond, silicon carbide provides a promising interface for spin and photons, even at ambient temperatures. Optically detected magnetic resonance (ODMR) and efficient spin control\cite{Widmann2015} have been demonstrated for both silicon vacancies and divacancies in SiC, using a combination of optical pulses for initialisation and readout, along with microwave (or radio frequency) pulses to manipulate the spin. Spin-photon entanglement with a fidelity of up to 76\% was recently demonstrated\cite{Fang2024}. From a spin coherence perspective, the inhomogeneous spin coherence time $T_2^*$ of divacancy centres lies on the microsecond time scale; and the Hahn echo results in a $T_2$ time of over 1 ms\cite{Christle2014} at cryogenic temperatures, without the need for isotropic purification. This $T_2$ coherence time can be further extended to over 5 s by dynamic decoupling and isotropic purification, utilising single-shot readout\cite{Anderson2022}. For silicon-vacancy centres, a $T_2^*$ of $\sim 30~\mu$s was recorded\cite{Nagy2019}, along with an Hahn-echo $T_2$ of around 1 ms\cite{Simin2017,Nagy2019,Babin2022}. The extension of $T_2$ on V2 centres has so far reached around 20 ms\cite{Simin2017}.\\

\noindent\textbf{Integration}\\
\noindent 
With its wide transparency window, concurrent second- and third-order optical nonlinearities, and compatibility with CMOS processing, SiC is emerging as a highly promising material for photonic and quantum applications. Recent advances in thin-film SiC have enabled its integration onto insulator substrates, leading to a wafer-scale platform known as silicon carbide on insulators (SiCOI)\cite{Lukin2020c, Song2019}. This advancement supports high-quality nanophotonic fabrication, achieving Q-factors up to $7\times10^6$ in micro-resonators (Fig.\ \ref{fig:fig4s}l)\cite{Guidry2020, Wang2021} and around $1\times10^6$ in photonic crystal cavities\cite{Song2019}. These structures enable coupling to both ensemble and single defect centres, facilitating Purcell enhancement\cite{Lukin2020c}, and laying the groundwork for integrated light-matter interaction platforms\cite{Hu2024}. While still in its early stages, SiCOI shows potential for hosting defect-centre devices without the need for hybrid integration. The platform already deliver essential functionalities such as low-loss propagation, high-speed switching\cite{Powell2022}, and frequency conversion\cite{Lukin2020c}, although these have yet to be demonstrated in conjunction with defect centres.\\

\section{Quantum emitters in quantum photonic applications}
\label{sec:chap4}
This chapter provides a brief overview of applications in quantum communications, computation, and sensing, highlighting the features and advantages of quantum emitters in each field.

\begin{table*}[!ht]
\centering
\renewcommand{\arraystretch}{3.2}    
\begin{center}
\resizebox{1\textwidth}{!}{
\begin{tabular}{|c|c|c|c|c|}
\hline
\multicolumn{2}{|c|}{\large \textbf{Material Platforms}}& \large \textbf{Quantum Dots} & \large \textbf{Diamond} & \large \textbf{Silicon Carbide}\\
\hline
\multicolumn{2}{|c|}{\large \textbf{Emission Wavelengths}} 
& \makecell{\textbf{GaAs/AlGaAs QDs: 775-800 nm} \\ \textbf{InGaAs/GaAs QDs: 900-950 nm} \\ \textbf{InAs/InP QDs: $\sim$ 1550 nm}} 
& \makecell{\textbf{NV$^-$: 637 nm} \\ \textbf{SiV$^-$: 737 nm} \\ \textbf{GeV$^-$: 603 nm} \\ \textbf{SnV$^-$: 619 nm}} 
& \makecell{\textbf{$\textrm{V}_{\textrm{Si}}^-$(V1): $\sim$ 862 nm} \\ \textbf{$\textrm{V}_{\textrm{Si}}^-$(V2): $\sim$ 917 nm} \\ \textbf{divacancies: $\sim$ 1100 nm}} \\
\hline
\multicolumn{2}{|c|}{\large \textbf{ZPL Ratio}} 
& \textbf{$>$ 95\%\textsuperscript{\cite{Somaschi2016}}} 
& \makecell{\textbf{NV$^-$: 3\%\textsuperscript{\cite{Santori2010}}} \\ \textbf{SiV$^-$: $\sim$ 70\%\textsuperscript{\cite{Tzeng2024}}} \\ \textbf{GeV$^-$: $\sim$ 60\%\textsuperscript{\cite{Siyushev2017}}} \\ \textbf{SnV$^-$: $\sim$ 60\%\textsuperscript{\cite{Gorlitz2020}}}} 
& \makecell{\textbf{$\textrm{V}_{\textrm{Si}}^-$(V1): 40\%\textsuperscript{\cite{Nagy2018}}} \\ \textbf{$\textrm{V}_{\textrm{Si}}^-$(V2): $<$ 10$\%$} \\ \textbf{divacancies: $<$ 10$\%$}} \\
\hline
\multicolumn{2}{|c|}{\large \textbf{Brightness}}
& \makecell{\textbf{Collection efficiency: 71$\%$\textsuperscript{\cite{Ding2025}}} \\ (laser with 80 MHz repetition rate \\ yields \textbf{57 MHz} count rates)} 
& \makecell{\textbf{Hundreds of kHz (CW)} \\ \textbf{\textsuperscript{\cite{Herrmann2024, Yurgens2024}}}} 
& \makecell{\textbf{Hundreds of kHz (CW)} \\ \textbf{\textsuperscript{\cite{Li2021, Castelletto2014}}}}  \\
\hline
\multirow{2}{*}{
\makecell{\large\textbf{Single}\\[2pt]\large\textbf{Photons}}
} & \large \textbf{Purity} & \textbf{$\sim$ 98$\%-$99$\%$ (pulsed)\textsuperscript{\cite{Scholl2019,Schweickert2018,Hanschke2018}}} 
& \makecell{\textbf{$\sim$ 97$\%-$99$\%$ (CW)\textsuperscript{\cite{Wang2022, Chen2022}}} \\ \textbf{$\sim$ 90$\%-$95$\%$ (pulsed)\textsuperscript{\cite{Arjona2022, Bersin2024}}}} 
& \makecell{\textbf{$\sim$ 95$\%$ (CW)\textsuperscript{\cite{Christle2017, Li2021}}} \\ \textbf{$\sim$ 88$\%$ (pulsed)\textsuperscript{\cite{Morioka2020}}}} \\   \cline{2-5}
&\makecell{\large \textbf{Indistinguish-} \\ \large \textbf{ability}}& \textbf{$\sim$ 98$\%$ (no filter)\textsuperscript{\cite{Zhai2022}}} 
& \textbf{$\sim$ 90$\%$ (filtered)\textsuperscript{\cite{Pompili2021, Bersin2024}}} 
& \textbf{$\sim$ 90$\%$ (filtered)\textsuperscript{\cite{Morioka2020}}} \\
\hline
\makecell[c]{\large\textbf{Entangled}\\[2pt]\large\textbf{Photons}} & \makecell[c]{\large \textbf{Entanglement} \\[2pt] \large \textbf{Fidelity}}& \makecell{\textbf{$\sim$ 98$\%$ (photon-photon)\textsuperscript{\cite{Huber2018}}} \\ \textbf{$>$ 80$\%$ (spin-photon)\textsuperscript{\cite{DeGreve2012, Coste2023}}}} & \makecell{\textbf{$\sim$ 91$\%$ (spin-photon)\textsuperscript{\cite{Stas2022}}} \\ \textbf{$\sim$ 85$\%$ (photon-nucleus)\textsuperscript{\cite{Stas2022}}}} & \textbf{$\sim$ 76$\%$ (spin-photon)\textsuperscript{\cite{Fang2024}}} \\
\hline
\multirow{2}{*}{ \makecell{\large \textbf{Spin}\\[2pt]\large\textbf{Coherence}}} & \large \textbf{$T_2$} & \textbf{GaAs/AlGaAs: 0.11 ms\textsuperscript{\cite{Zaporski2023}}} 
& \makecell{\textbf{NV$^-$: $\sim$ 1 s (DD)\textsuperscript{\cite{Abobeih2018}}} \\ \textbf{Group IV: $\sim$ 1 ms - 10 ms (DD)} \\ \textbf{\textsuperscript{\cite{Guo2023, Sukachev2017, Senkalla2024}}}} 
& \makecell{\textbf{$\textrm{V}_{\textrm{Si}}^-$(V1): $\sim$ 1 ms (Hahn)\textsuperscript{\cite{Babin2022, Nagy2019, Simin2017}}} \\ \textbf{$\textrm{V}_{\textrm{Si}}^-$(V2): $\sim$ 20 ms (Hahn)\textsuperscript{\cite{Simin2017}}} \\ \textbf{divacancies: $\sim$ 5 s (DD)\textsuperscript{\cite{Anderson2022}}}} \\
\cline{2-5}
& \large \textbf{$T_2^*$}
& \makecell{\textbf{GaAs/AlGaAs: 0.6 $\mu$s\textsuperscript{\cite{Nguyen2023}}} \\ \textbf{InGaAs/GaAs: 125 ns\textsuperscript{\cite{Jackson2022}}}}
& \makecell{\textbf{NV$^-$: $\sim$ 1.5 ms\textsuperscript{\cite{Herbschleb2019}}} \\ \textbf{Group IV: $\sim$ 1 $\mu$s $-$ 10 $\mu$s\textsuperscript{\cite{Sohn2018, Senkalla2024}}}}
& \makecell{\textbf{$\textrm{V}_{\textrm{Si}}^-$(V1): $\sim$ 30 $\mu$s\textsuperscript{\cite{Nagy2019}}} \\ \textbf{divacancies: $\sim$ 1 $\mu$s\textsuperscript{\cite{Christle2014}}}} \\
\hline
\end{tabular}}
\end{center}
    \caption{\textbf{Summary of quantum and photonic properties across different material platforms.}}
    \label{table:table1}
\end{table*}

\subsection{Quantum Communications}

Empowered by the `no-cloning theorem' and quantum properties such as superposition and entanglement, quantum communications provide security based on the laws of quantum mechanics. One prominent application, quantum key distribution (QKD), encrypts cryptographic keys into the quantum states of photons, such as polarisation, phase, quadratures, or photon numbers. Among various QKD protocols\cite{Minder2019,Cao2020,Zhang2022b,Zhou2023b}, attenuated coherent laser pulses (decoy states) are widely implemented as the photonic resource. While the original QKD protocol, developed by Bennett and Brassard in 1984 (BB84), relied on a perfect single-photon source, the use of decoy states initially arose as a compromise due to the lack of a reliable single-photon source. Over time, however, this approach has been driven by both the cost-effectiveness and the simplicity of its implementation.

\begin{figure*}[ht!]
    \centering
    \includegraphics[width = 15 cm]{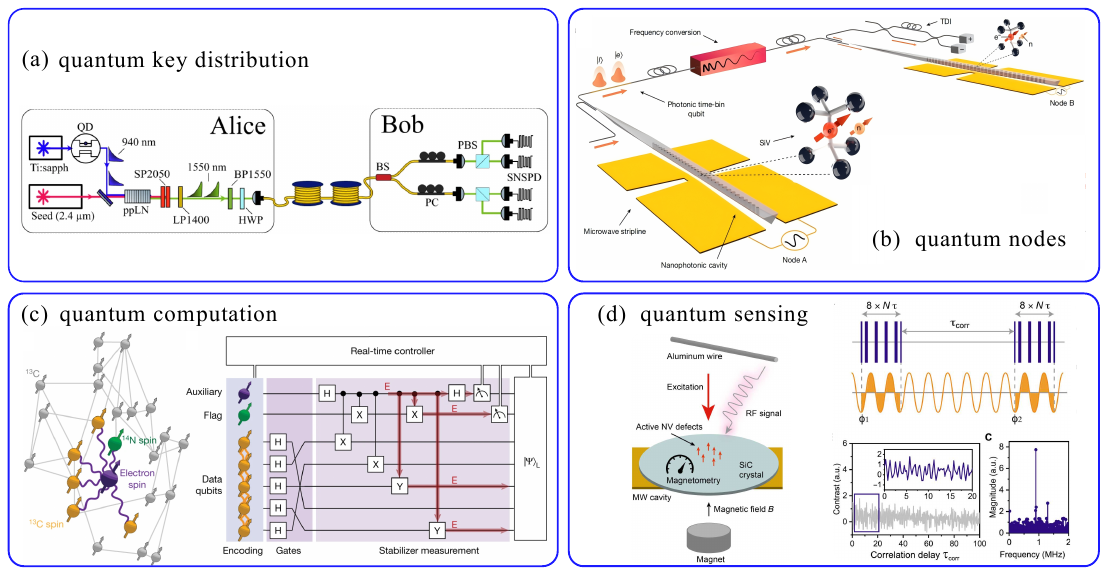}
    \caption{\textbf{Applications of quantum emitters in photonic quantum technologies.} (a) Quantum key distribution setup with a QD-based single-photon source. (b) Two-node quantum network using SiV centres in diamond cavities. (c) Qubit generation with NV centres and $^{13}$C nuclear spins in diamond. (d) Quantum sensing of radio-frequency signals using defect centres in SiC. Panels reproduced with permission from: (a) Ref. \cite{Morrison2023}; (b) Ref. \cite{Knaut2024}; (c) Ref. \cite{Abobeih2022}; (d) Ref. \cite{Jiang2023}, under a Creative Commons license CC BY 4.0.    }
    \label{fig:fig5}
\end{figure*}

With the increasing maturity of QD-based single-photon sources, a growing number of QKD trials involving single photons have been demonstrated (see Fig.\ \ref{fig:fig5}a)\cite{Gao2022,Yang2024,Zhang2024}. Secure key rates of $5.6\times 10^{-2}$ in the lab and $1.1\times 10^{-3}$ over free space have been achieved -- both exceeding the fundamental upper bound of those based on attenuated lasers under the same environmental conditions\cite{Zhang2024}, demonstrating the advantages of true single-photon sources. Furthermore, entangled photon pairs generated by QDs have been employed in entanglement-based protocols, such as BBM92\cite{Schimpf2021b}, E91\cite{Basset2021}, with entanglement swapping (a photonic version) and quantum teleportation also demonstrated\cite{Zopf2019,Basset2019,Anderson2020}. These bright single-photon and entangled-photon resources hold great promise for advancing the development of device-independent QKD, where weak coherent lasers do not work.

Going beyond QKD, future large-scale quantum networks are envisioned to comprise quantum channels and nodes (Fig.\ \ref{fig:fig5}b). Each node must be capable of generating, processing, and temporarily storing quantum information (typically carried by photons). A coherent light-matter interface is crucial for constructing these quantum nodes. To date, multi-node quantum networks have been demonstrated using both NV centres \cite{Kalb2017, Pompili2021, Hermans2022} and SiV centres in diamond\cite{Stas2022, Bersin2024, Knaut2024}, where each node consists of an electron qubit and a nuclear qubit -- the latter serving as a memory while the former mediates interactions between photons and nuclear-spin quantum memory. These nodes have enabled advanced networking functionalities, including quantum teleportation \cite{Hermans2022}, entanglement distillation \cite{Kalb2017}, and entanglement between distant quantum memories. Additionally, the photonic properties of these nodes have been improved by integrating them into diamond cavities to enhance brightness\cite{Reindl2017,Yurgens2024}, and by wavelength conversion to telecom wavelengths to reduce losses in fibre transmission\cite{Stolk2022}. Other platforms, including atoms \cite{Jing2019,Liu2024, Liu2024b}, rare-earth-doped solids \cite{Lago2021}, and epitaxial QDs\cite{Delteil2016,Yu2023}, have also made promising progress in enabling quantum node functionality.

\subsection{Quantum Information processing}

Multiple physical platforms have been pursued for universal quantum computation, including superconducting qubits, Rydberg atoms, trapped ions, spin qubits, and photons. Diamond has also emerged as a contender, with multi-qubit processors predominantly based on NV centres. These systems couple a central electron spin to surrounding $^{13}$C nuclear spins \cite{Bradley2019, Abobeih2022} (see Fig.\ \ref{fig:fig5}c). Recent demonstrations have shown high-fidelity single- and two-qubit gates \cite{Sekiguchi2022,Xie2023}, minute-scale coherence times \cite{Bradley2019}, and fault-tolerant operations on a five-qubit error correction code \cite{Abobeih2022}. However, scaling the qubit may lead to crosstalk and selective control challenges. A modular approach, such as optical coupling between different defect centres, could help address these limitations. Beyond NV centres, other defect centres in diamond, defects in SiC, and QDs are also promising. Notably, a recent demonstration employed a central electron spin surrounded by $\sim 10^5$ nuclear spins in low-noise GaAs QDs to form a quantum register\cite{Appel2025}.

Quantum emitters also play a key role in photonic quantum information processing. Due to weak photon-photon interactions, photonic quantum computation primarily relies on linear optics and quantum teleportation. Specialised protocols, such as boson sampling, which demonstrates quantum advantage in simulating bosonic interference\cite{Loredo2017,Wang2019a}, and variational quantum eigensolvers\cite{Maring2024}, used for approximating ground states in quantum chemistry, have been actively explored. A scalable approach toward universal quantum computation is measurement-based quantum computation (MBQC), where photonic qubits are arranged in a large-scale entanglement network and subjected to projective measurements, with each measurement outcome dictating subsequent operations through feedforward control.

A major challenge in MBQC is preparing high-dimensional, large-scale entanglement networks. The current approach follows this strategy: First, small constant-size entangled qubits, such as few-photon cluster states, are generated simultaneously using arrays of quantum light sources. These entangled qubits, considered the basic resource, are then interconnected through projective operations, known as fusion gates, gradually forming a larger entanglement network\cite{Bartolucci2023}. As quantum projective measurements are performed on the network, entanglement is consumed, reducing the network size. Meanwhile, quantum light sources and fusion gates replenish the entanglement, ensuring the continuous operation of the computation.

With a bright and coherent spin-photon interface and excellent photonic properties, QDs can serve as deterministic resource state generators, producing small constant-size entangled qubits. Using a time-bin entanglement protocol \cite{Appel2022} and leveraging the spin-photon interface, few-photon cluster states and GHZ states have been deterministically generated \cite{Meng2024}. Alternatively, few-photon cluster states can be deterministically generated from QDs via precise timing of the excitation pulse \cite{Cogan2023}, or by utilising the high brightness of QD-based single-photon sources and probabilistic entangling gates \cite{Istrati2020}.

\subsection{Quantum Sensing}
Quantum sensing leverages quantum objects or properties, such as quantum coherence, correlation, and entanglement, to measure physical quantities \cite{Degen2017}. Spin defects in diamond and SiC are exceptional quantum sensors, hosting single or ensemble spins with long coherence times. These spins are easily manipulated using microwave control and function across a broad temperature range, from cryogenic to room temperature. Defect-spin-based magnetometry has been successfully demonstrated in both diamond and SiC \cite{Abraham2021, Xie2021} (Fig.\ \ref{fig:fig5}d), with NV centres making significant strides toward real-world applications and commercialisation. Sensitivities on the order of nT/Hz$^{-1/2}$, and even down to sub-pT/Hz$^{-1/2}$, are achievable using NV ensembles at room temperature \cite{Xie2021}. Moreover, single NV spins have become an increasingly powerful and versatile tool\cite{Rovny2024} for high-precision, atomic-scale probes in condensed matter and bio-sensing. For example, spatially resolved magnetism in 2D van der Waals magnets is mapped using a single NV centre integrated into a nanotip\cite{Thiel2019}. Beyond magnetometry, defect centres in diamond and SiC are useful for high-sensitivity detection of temperature\cite{Zhou2017}, electric fields\cite{Dolde2011}, and strain fields\cite{Falk2014}.

Apart from spin-based quantum sensing, photon-based sensing and metrology leveraging quantum properties offer precision beyond the classical limit \cite{Pirandola2018, Couteau2023}. QDs are versatile quantum light sources with the ability to generate single photons, multiphoton entanglement \cite{Istrati2020}, Schrödinger-cat states \cite{Cosacchi2021}, and intensity squeezing \cite{Iles2019}. Due to their high brightness and advanced photonic state engineering capabilities, QD-based quantum light sources hold significant promise for discrete-variable photonic quantum sensing.

\section{Conclusion}
\label{sec:chap5}
To summarise, we have presented a set of metrics for evaluating the performance of quantum emitters in quantum applications. A brief overview of three types of quantum emitters -- epitaxial quantum dots, defect centres in diamond, and defect centres in SiC -- is provided, covering the preparation methods, energy level structures, optical properties, and integration feasibility. Additionally, we analysed their applications across various quantum technologies, with a focus on recent advancements.

Comparing the three material platforms (Table\ \ref{table:table1}): epitaxial QDs exhibit excellent optical properties -- coherent, bright, and \textit{on demand} -- but have relatively limited coherence, both for electron spins and nuclear spins, compared to diamond and SiC. This relatively limited coherence (on the $\mu$s to ms timescale) arises from the presence of a dense nuclear spin ensemble surrounding the trapped electron spin. The electron coherence in QDs may limit their spin-photon applications in long-distance communication protocols. However, spin control in QD systems is much faster than the commonly used microwave drives in diamond and SiC, which can be beneficial for resource state generation in MBQC, where QDs create few-photon cluster states.

Defect centres in diamond, in contrast, offer significant advantages in spin coherence, particularly for NV$^-$ centres, which maintain long coherence even at elevated temperatures. Despite the narrow linewidth (excluding spectral diffusion) and near-zero $g^{(2)}(0)$, the coherent part of emitted photons is significantly hindered by a large phonon sideband component. As a result, high photon indistinguishability is only achieved after tight spectral filtering or temporal gating. This spectral or temporal filtering leads to a high probability of photon removal, which can be effectively viewed as photon loss or an \textit{erasure error}. Moreover, existing diamond defect centres all emit in the red, well outside the low-loss window for fibre-optic communication.

Similar to diamond, defect centres in SiC also excel in their spin properties, enabling efficient spin-photon interfaces at room temperature. Some defects emit near the telecom O-band, offering more favourable wavelengths for photonic application. However, they still face photonic challenges akin to those in diamond, and the spin-photon interface, including entanglement between electron and nuclear spins, remains less advanced. 

From an integration perspective, all three material systems are still in the early stages of scaling towards large-scale integrated photonics. Silicon carbide on insulators is a promising platform with relatively strong $\chi^{(2)}$- and $\chi^{(3)}$-nonlinearity, as well as a broad transparency window and low loss\cite{Yi2022, Lukin2020}. III-V semiconductor membranes, such as GaAs and AlGaAs, which host QDs, also exhibit both strong $\chi^{(2)}$- and $\chi^{(3)}$-nonlinearity\cite{Chen2024c}, but their losses are slightly higher, especially when vertical n-i-p diodes are incorporated. Diamond membranes, on the other hand, exhibit only $\chi^{(3)}$ nonlinearity\cite{Hausmann2014}, which limits the implementation of high-speed electro-optical switches that rely on the Pockels effect.

Looking ahead, quantum emitters must improve in several aspects to meet the demands of scalable quantum technologies. QDs must be integrated with a quantum memory that efficiently exchanges information with the QD spin qubit. Encouraging progress has been made in realising prototype memory functions in the proximal nuclear-spin ensemble \cite{Appel2025, Lei2025}. Additionally, photonic states can be stored in atomic memories \cite{Jing2024}, such as those based on rubidium and caesium vapours \cite{Mottola2023, Maa2025}. Moreover, while precise pre-location techniques have been developed, achieving site-controlled QD growth would greatly benefit large-scale nanophotonic fabrication, enabling dense arrays of light-matter interfaces. Colour centres in diamond and SiC should be coupled to high-finesse optical cavities to enhance the collection efficiency of the ZPL while suppressing emission into the phonon sideband. Preliminary results using microcavities show finesse values of $10^4$ and $10^3$ at cryogenic temperatures for NV and SnV centres in diamond membranes\cite{Yurgens2024, Herrmann2024}, respectively. The integration of defects into high-finesse cavities remains elusive, although the field is rapidly advancing.

Other common challenges include frequency conversion or the development of high-quality quantum emitters directly at telecom wavelengths. Hetero-integration into versatile integrated platforms, such as silicon nitride and lithium niobate, is also a highly pursued direction. Beyond the emitters introduced here, the emergence of several other emitter platforms, including single rare-earth ions in solids\cite{Kindem2020}, defects in silicon\cite{Higginbottom2022}, and emitters in two-dimensional materials like wide-bandgap hexagonal boron nitride and transition metal dichalcogenides\cite{Stern2024}, presents new opportunities for the field. Quantum emitters are thus becoming a flourishing avenue for quantum photonics research.

\section*{Competing interests}
The authors declare no competing interests.\\

\section*{Acknowledgements}
We gratefully acknowledge the support from the Sichuan Science and Technology Programme (2024ZYD0050, 2024YFHZ0369).


\end{document}